# First-Principles Thermodynamic Theory of Seebeck Coefficients

Yi Wang[+], Yong-Jie Hu, Shun-Li Shang, Bi-Cheng Zhou, Zi-Kui Liu, and Long-Qing Chen*

Department of Materials Science and Engineering, The Pennsylvania State University,

University Park, PA 16802, USA

Thermoelectric effects, measured by the Seebeck coefficients, refer to the phenomena in which a temperature difference or gradient imposed across a thermoelectric material induces an electrical potential difference or gradient, and vice versa, enabling the direct conversion of thermal and electric energies. All existing understanding and first-principles calculations of Seebeck coefficients have been based on the Boltzmann kinetic transport theory. Here we demonstrate that the Seebeck coefficient is a well-defined thermodynamic quantity that can be determined from the change in the chemical potential of electrons induced by the temperature change and thus can be efficiently computed solely based on the electronic density of states through first-principles calculations at different temperatures. The proposed approach is demonstrated using the prototype PbTe and SnSe thermoelectric materials. The proposed thermodynamic approach dramatically simplifies the calculations of Seebeck coefficients, making it possible to search for high performance thermoelectric materials using high-throughput first-principles calculations.

[+]**yuw3@psu.edu**; *lqc3@psu.edu

# I. INTRODUCTION

Thermoelectric effects refer to the phenomena that a temperature gradient induces an electromotive force across a thermoelectric material and *vice versa*, enabling the direct conversion of thermal and electric energies and therefore drawing unprecedented interests in the field of clean energy technologies.[1–4] There have been extensive efforts in the search for thermoelectric materials with the best performances quantitatively measured by the figure of merit [5,6]

$$\text{Eq. 1} \quad ZT = \frac{\sigma \alpha^2}{\kappa} T$$

where $\sigma$ is the electric conductivity, $\kappa$ is the thermal conductivity, and $T$ is temperature. The Seebeck coefficient $\alpha$ in Eq. 1 measures the magnitude of the thermoelectric effect and is given by ratio of the induced electromotive force ($\Delta \phi$) to the temperature difference ($\Delta T$) across a thermoelectric material,[7] i.e.

$$\text{Eq. 2} \quad \alpha = -\left(\frac{\Delta \phi}{\Delta T}\right)_p$$

where the subscript $p$ represents the pressure at which the thermoelectric coefficient is measured at zero current density. When the temperature difference becomes infinitesimally small, the Seebeck coefficient becomes

$$\text{Eq. 3} \quad \alpha = -\left(\frac{\partial \phi}{\partial T}\right)_p.$$

While $S$ and $K$ in Eq. 1 are kinetic coefficients, Eq. 3 shows that the Seebeck coefficient is a well-defined thermodynamic property representing the derivative of electric potential ($f$) with respect to thermal potential ($T$).

However, it is not always clear in the literature whether the Seebeck coefficient can be defined purely from thermodynamics. [8] In the existing literature, the kinetic Boltzmann transport theory [9–11] is typically employed for determining the thermoelectric effects from first-principles calculations. For instance, in the Cutler-Mott theory [9] the Seebeck coefficient is computed as

$$\text{Eq. 4} \quad \alpha = \frac{1}{qT} \frac{\int [\varepsilon - \mu] \sigma(\varepsilon) d\varepsilon}{\int \sigma(\varepsilon) d\varepsilon}$$

where $q = \pm e$ with the "–" sign being for electrons as charge carriers, the "+" sign being for holes, and "$e$" being the amount of elementary charge, $\varepsilon$ represents the electronic band energy, $\mu$ the Fermi level, and $\sigma(\varepsilon)$ is a kinetic coefficient called the energy-dependent differential electrical conductivity. The first-principles implementations of the Cutler-Mott theory rely on a further assumption on the mechanism of electron scattering, e.g., the constant relaxation time approximation.

In this work, we present a purely thermodynamic approach to computing the Seebeck coefficient. Such an approach dramatically simplifies the first-principles calculations of Seebeck coefficients, making it possible to perform the calculations in a high-throughput manner and thus an efficient first-principles search for thermoelectric materials with high Seebeck coefficients.

## II. THERMODYNAMIC THEORY OF THE SEEBECK COEFFICIENT

There are a number of existing attempts to define the Seebeck coefficient based on thermodynamics, and the earliest can be dated back to the work by Callen in 1948[7]. For example, Wood [12] defined the Seebeck coefficient as the total derivative of the electrochemical potential of electrons with respect to temperature, which was later called the Kelvin formula by Peterson and Shastry [13] while changing the total derivative into a partial derivative at constant volume. However, as pointed out by Apertet et al., [8] it is not clear whether the potential gradient in the Kelvin formula is the chemical potential or the electrochemical potential gradient. Furthermore, to our knowledge, all existing applications based on the Kelvin formula have been limited to model systems. [8,13,14]

Here we propose a relation between the electric potential $\phi$ in Eq. 3 and the chemical potential of electrons calculated using the Mermin's finite temperature extension of density functional theory (DFT) [15,16] under the framework of the modern first-principles approach. [17–22] According to Ziman [23], the *absolute thermal electric force*, $\Theta$, can be defined as

Eq. 5 $\quad \Theta(T) = \int_0^T \alpha dt$

Since there are no other mobile charged species in the system except the electronic charge carriers, the difference in chemical potential of electrons due to the temperature difference must be the only thermodynamic source responsible for the thermoelectric electromotive force. For a uniform material with uniform temperature, a temperature change results in the change in the chemical potential of thermal electrons, i.e., the temperature dependent portion of the free energy gain per electron. For a non-uniform system, each point in space can be considered as a thermodynamic system, and thus a system with a non-uniform temperature distribution has non-uniform distribution of chemical potentials of electrons. The gradient in chemical potential of electrons is the thermodynamic driving force.

By relating Ziman equation (Eq. 5) to the chemical potential of electrons, we propose to relate the absolute thermal electric force to the chemical potential of electrons,

Eq. 6 $\quad \phi(T) = \frac{1}{q}\int_0^T \frac{\partial \mu(t)}{\partial t} dt = \frac{1}{q}[\mu(T) - \varepsilon_F] = \frac{1}{q}\zeta(T)$

where $\mu(T)$ is the chemical potential of electrons (or Fermi level) by the Mermin's theory [15,16] and $\varepsilon_F$ is the Fermi energy. The notation of $\zeta = \mu - \varepsilon_F$ is a reminiscent of that given by Sommerfeld, [24] so that Eq. 6 sets $\phi = 0$ at 0K. Following the previous calculations of the thermal electronic contribution to thermodynamic properties at finite temperatures [25,26] based on the Mermin's theory, [15,16] we can obtain the temperature dependence of chemical potential of electrons based on the electron density of states.

Considering the fact that the electrons are explicitly treated in the current implementation [20,21] of DFT, hereafter in all the formulations we will use $q = -e$. To obtain the Seebeck coefficient under the Mermin's theory, we can start from the conservation equation

Eq. 7 $\quad \int n(\varepsilon, V) f d\varepsilon = N$

where $n(\varepsilon,V)$ is the electronic density of states (e-DOS), $N$ is the total number of electrons in the considered system, and $f$ is the Fermi-Dirac distribution

Eq. 8 $\quad f = \dfrac{1}{\exp\left[\dfrac{\varepsilon + e\phi}{k_B T}\right] + 1}$

Here we note that in Eq. 7 and Eq. 8, the Fermi energy $\varepsilon_F$ has been taken as the reference for the band energy $\varepsilon$.

Finding the full derivative of Eq. 7 under the isobaric condition and noting that $N$ is a constant and $V$ is temperature-dependent, we have

Eq. 9 $\quad \left(\dfrac{\partial V}{\partial T}\right)_P \int \dfrac{\partial n(\varepsilon,V)}{\partial V} f d\varepsilon + \int n(\varepsilon,V) \dfrac{\partial f}{\partial T} d\varepsilon = 0$

with

Eq. 10 $\quad \dfrac{\partial f}{\partial T} = \dfrac{1}{k_B} f(1-f) \left[ \dfrac{-e}{T}\left(\dfrac{\partial \phi}{\partial T}\right)_P + \dfrac{\varepsilon - \zeta}{T^2} \right]$

After a few rearrangements, we obtain the formulation to calculate the Seebeck coefficient under the isobaric condition, $\alpha_P$

Eq. 11 $\quad \alpha_P = -\left(\dfrac{\partial \phi}{\partial T}\right)_P = -\dfrac{k_B T \beta}{e\eta} \int \dfrac{\partial n(\varepsilon,V)}{\partial V} f d\varepsilon - \dfrac{1}{eT\eta V} \int n(\varepsilon,V) f(1-f)(\varepsilon - \zeta) d\varepsilon$

where $\beta$ is the volume thermal expansion coefficient and

Eq. 12 $\quad \eta = \dfrac{1}{V} \int n(\varepsilon,V) f(1-f) d\varepsilon$

From Eq. 11, it is seen that the constant pressure Seebeck coefficient contains two terms, one accounts for the thermal expansion contribution – the first term in the right hand side of Eq. 11, and the other accounts for the constant volume contribution – the second term in the right hand side of Eq. 11. Eq. 11 demonstrates that the Seebeck coefficient can be computed from the e-DOS instead of the electric conductivity as in the Boltzmann transport theory. [9,27] While e-DOS, $n(\varepsilon,V)$, is readily accessible from first-principles calculations, [28–30] $\sigma(\varepsilon)$ in Eq. 4 is a much more challenging quantity to compute, which requires the calculation of electron group velocity and the assumption of a constant relaxation time. [31] Therefore, the present formulation of Eq. 11 provides a very efficient computational method to search for superior thermoelectric materials employing high-throughput first-principles calculations.

Further output from the present approach is about the finite temperature calculation of mobile charge carrier concentration $\eta$ in Eq. 12. The integral $\int n(\varepsilon,V)f(1-f)d\varepsilon$ can be considered as the total number of mobile carriers in a thermoelectric solid – the number of electrons participating in the conduction process at finite temperature. This can be understood in the sense that the pair product of $f(1-f)$ represents the possibility that the electrons occupied "$f$ number" of electronic states with energy $\varepsilon$, transmitted to (or the vice versa), the "1-$f$ number" of unoccupied electronic states with energy $\varepsilon$ at finite temperature. Accordingly, $n(\varepsilon,V)f(1-f)$ represents the density of states of charge carriers at finite temperature.

## III. COMPARISON WITH THE BOLTZMANN TRANSPORT THEORY

### A. Low temperature limit

First, we can compare the present formulation with that derived from the kinetic Boltzmann transport theory [9–11] at the limit of $T \to 0$. Using Sommerfeld's low temperature expansion, [24,32] we get

Eq. 13 $\phi = -\dfrac{\pi^2}{6e}(k_B T)^2 \dfrac{\partial \ln n(\varepsilon,V)}{\partial \varepsilon}\bigg|_{\varepsilon=0}.$

Therefore at $T \to 0$, we have

Eq. 14 $\alpha_V = -\dfrac{\pi^2 k_B^2 T}{3e}\dfrac{\partial \ln n(\varepsilon,V)}{\partial \varepsilon}\bigg|_{\varepsilon=0}.$

This relation tells that at the limit of $T \to 0$, a thermoelectric material is *p*-type if its e-DOS has a negative slope at the Fermi energy with increasing band energy whereas it is *n*-type if its e-DOS has a positive slope at the Fermi energy with increasing band energy. This is the situation for most insulators around the top of the valence band and the bottom of the conduction band.

On the other hand, for the Boltzmann transport theory [9–11] at the limit of $T \to 0$ [33]

Eq. 15 $\alpha_V = -\dfrac{\pi^2 k_B^2 T}{3e}\left[\dfrac{\partial \ln \bar{n}(\varepsilon,V)}{\partial \varepsilon}\bigg|_{\varepsilon=\varepsilon_F} + \dfrac{\partial \ln v^2(\varepsilon)}{\partial \varepsilon}\bigg|_{\varepsilon=\varepsilon_F} + \dfrac{\partial \ln \tau(\varepsilon)}{\partial \varepsilon}\bigg|_{\varepsilon=\varepsilon_F}\right]$

where $\bar{n}(\varepsilon,V) = n(\varepsilon - \varepsilon_F, V)$, $v$ is an average electron velocity, and $\tau$ is the so-called relaxation time. In recent first-first-principles calculations, [11,27,34–38] $\tau$ is mostly assumed to be constant. As pointed by MacDonald, [33] the constant $\tau$ approximation is a rather rude one at $T \to 0$. As a matter of fact, after using a correction for the free electrons with "screening" charge together with a constant mean free path, instead of constant relaxation time, Wilson (see the book by MacDonald [33]) found that at $T \to 0$

Eq. 16 $\alpha_V = -\dfrac{\pi^2 k_B^2 T}{3e}\dfrac{1}{\varepsilon_F}$

with the parabolic band approximation [36] that $\bar{n}(\varepsilon) \propto \varepsilon^{1/2}$ and $v^2(\varepsilon) \propto \varepsilon$. Note that the evaluation of $\bar{n}(\varepsilon,V)$ in Eq. 15 and Eq. 16 at $\varepsilon = \varepsilon_F$ is equivalent to evaluating $n(\varepsilon,V)$ in Eq. 14 at $\varepsilon = 0$.

### B. Relation between the present formulation and the Boltzmann transport theory

By the Boltzmann transport theory, [10] $\sigma(\varepsilon)$ in Eq. 4 can be calculated by

$$\text{Eq. 17} \quad \sigma(\varepsilon) = e^2 f(1-f) \sum_{\mathbf{k}} \mathbf{v_k} \mathbf{v_k} \tau_{\mathbf{k}}$$

where $\mathbf{v_k}$ is the group velocity of the charge carriers with the crystal momentum $\mathbf{k}$. To relate the result from Eq. 4 using the Boltzmann transport theory to that from Eq. 11 (neglecting the effect of thermal expansion) using the present formulation, we can introduce an approximation by Heisenberg uncertainty principle, such that $\mathbf{v_k}$ and $\tau$ in Eq. 15 approximately satisfy

$$\text{Eq. 18} \quad \mathbf{v_k} \mathbf{v_k} \tau = \frac{m_e \mathbf{v_k}}{m_e} \cdot \mathbf{v_k} \tau \propto \frac{\Delta \mathbf{p} \Delta \mathbf{l}}{m_e} \propto \frac{\hbar}{2m_e} f(\frac{T_0}{T}).$$

where $m_e$ is the mass of electrons, $\hbar$ is the reduced Planck constant, and $f$ only depends on temperature.

### C. Incorporating the effect of thermal expansion to the Boltzmann transport theory

In case that one insists using the Boltzmann transport theory to calculate the Seebeck coefficient, the effect of thermal expansion can be included by

$$\text{Eq. 19} \quad \alpha_p = -\frac{k_B T \beta}{e\eta} \int \frac{\partial n(\varepsilon, V)}{\partial V} f d\varepsilon - \frac{1}{eT} \frac{\int [\varepsilon - \zeta] f(1-f) \sum_{\mathbf{k}} \mathbf{v_k v_k} \tau_{\mathbf{k}} d\varepsilon}{\int f(1-f) \sum_{\mathbf{k}} \mathbf{v_k v_k} \tau_{\mathbf{k}} d\varepsilon}$$

Here note that the reference point of the band energy $\varepsilon$ have been shifted by treating the Fermi energy as reference.

## IV. COMPUATIONAL DETAILS

The actual validation of the proposed formalism is carried out by performing first-principles calculations for the widely studied thermoelectric material PbTe [39,40] and the new thermoelectric material SnSe discovered by Zhao et al. [41–43] For simplicity, we assume the rigid band approximation, [10] i.e., the band structure is assumed to remain unchanged from doping or from the thermal electronic excitation at finite temperatures. Therefore, a positive doping (p-doping or hole doping) only shifts the Fermi energy toward a low energy, transforming the insulator into a p-type conductor even at 0K. Similarly, a negative doping (n-doping or electron doping) increases the Fermi energy above the (bottom) edge of the conduction band, transforming the insulator into an n-type conductor even at 0K.

Here we present details for the first-principles calculations. One needs to first overcome several issues for the first-principles calculations. The major issue is concerned with the band gap which is mostly underestimated by the commonly employed Generalized gradient approximation [44] and local density approximation.[45] Fortunately for PbTe, the pioneering calculations performed by Singh [46] showed that the Engel-Vosko generalized gradient approximation [47] plus spin-orbit coupling as implemented in WIEN2k package [22] gave an excellent band structure and band gap. Therefore, we followed Singh's work for the e-DOS calculation for both PbTe and SnSe. For the e-DOS calculation of PbTe using WIEN2k, we follow exactly the same settings by Singh, i.e., linearized augmented plane-wave sphere radii ($R$) of 2.9 Bohr were used for both Pb and Te; $R \times k_{max}$=9.0 where $k_{max}$ is the interstitial plane-wave cutoff; and 48×48×48 $k$-mesh. These settings, together with the Engel-Vosko generalized gradient approximation plus spin-orbit coupling, are the key to producing the band gap and e-DOS feature in the range -0.35 ~ 0.0 eV as shown in Figure 1. For the e-DOS calculation of SnSe using WIEN2k, linearized augmented plane-wave sphere radii ($R$) of 2.5 Bohr was used for both Sn and Se together with $R \times k_{max}$=7.0 and 28×26×10 $k$-mesh.

The second issue is concerned with the calculations of the lattice parameter as a function of temperature and the resulted thermal expansion. The good e-DOS from the Engel-Vosko generalized gradient approximation is at the cost of losing accuracy of the total energy. We overcame the problem by invoking the Perdew-Burke-Ernzerhof revised for solids (PBEsol) [48] exchange-correlational functional as implemented in the Vienna *ab initio* simulation package (VASP, version 5.3) [20,21]. For PbTe, the Pb $5d^{10}6s^26p^2$ and Te $6s^26p^4$ electrons have been treated in the valence states; An energy cutoff of 336.7 eV and 20×20×20 $k$-mesh was used for calculating the total energy. At room temperature, aided by the quasiharmonic phonon approximation, PBEsol provides a lattice parameter of 0.3242 nm which is within 0.3% of the experimental value of 0.3232 nm [11] for PbTe. For SnSe, the Sn $4d^{10}5s^25p^2$ and Se $5s^25p^4$ electrons were treated in the valence states; An energy cutoff of 336.7 eV and 18×17×7 $k$-mesh was used for calculating the total energy. At room temperature, using the quasiharmonic phonon approximation, PBEsol results in lattice parameters of 0.4159, 0.4397, and 1.1469 nm which are comparable to the experimental values of 0.4153, 0.4445, and 1.1501 nm [49] for SnSe.

The third issue is about the phonon calculation for polar insulators in order to obtain the finite temperature thermodynamic properties. For this issue, the present authors have proposed a mixed space approach [50–53] to account for the dipole-dipole interactions for a phonon calculation in the real space using supercell method. The required inputs of Born effective charge and dielectric constant tensors to the mixed-space approach are calculated following the linear-response approach by Gajdoš et al. [54]. For PbTe, an energy cutoff of 237.8 eV and 3×3×3 $k$-mesh and a 4×4×4 supercell containing 128 atoms were used for phonon calculation. For SnSe, an energy cutoff of 241.1 eV and 3×3×3 $k$-mesh and a 2×2×2 supercell containing 54 atoms were used for phonon calculations.

## V.  NUMERICAL RESULTS AND DISCUSSIONS

Figure 1 illustrates the evolution of electron density of states and the corresponding chemical potential as a function of temperature in single crystal PbTe.  PbTe is an intrinsic semiconductor as indicated by its 0K e-DOS shown in Figure 1a. Therefore, at 0K, PbTe is an insulator since the conduction band is unoccupied and separated by an energy gap from the completely filled valence band.  As temperature increases, the electron occupation among electronic states changes. When the e-DOS curve has a negative slope at the 0 K Fermi energy with respect to the band energy as in the case of p-type PbTe shown in Figure 1a, i.e. there are fewer states for the electrons to occupy with increasing energy, the chemical potential of electrons will increase with increasing temperature as shown in Figure 1b. Similarly, when the e-DOS curve has a positive slope with increasing energy at the 0 K Fermi energy with respect to the band energy, i.e. there are more states for the electrons to occupy with increasing energy as in the case of n-type PbTe Figure 1a, the chemical potential of electrons will decrease with increasing temperature as shown in Figure 1c.

Figure 2 compares the calculated Seebeck coefficients with measurements for PbTe at a variety of p- and n-doping levels. The minor difference between the calculated and measured data for the p-type doping levels of $2.0\times10^{17}$ cm$^{-3}$ could be due to the experimental uncertainty, while the slight difference between the calculated and measured data for the n-type doping levels of $1.4\times10^{20}$ cm$^{-3}$ could be due to the rigid band approximation at high doping levels.

Figure 3 shows a comparison of the present approach with those calculated using BoltzTrap [11] by Madsen and Singh based on the Boltzmann transport theory, at the same p-type doping levels of $1.0\times10^{18}$, $5.0\times10^{18}$, and $1.1\times10^{19}$ cm$^{-3}$, and n-type doping levels of $1.1\times10^{19}$, $5.0\times10^{18}$, and $1.0\times10^{18}$ cm$^{-3}$. To support the comparison, also plotted Figure 3 are the available experimentally measured data with similar carrier concentrations at p-type doping levels of $5.3\times10^{19}$ cm$^{-3}$ by Heremans and coauthors [21,22] and n-type doping levels of $1.0\times10^{19}$ and $5.8\times10^{18}$ cm$^{-3}$ by LaLonde et al. [55]. It is indeed observed that the present efficient thermodynamic approach produces very similar results with the time-consuming kinetic approach by Singh [46]. In the moderate temperature range, the calculated Seebeck coefficients by the present approach have larger slopes than those calculated by Singh using BoltzTrap [11]. This is likely due to the fact that the thermal expansion effects were not accounted for in the BoltzTrap code. The same is observed for the calculated results for SnSe in Figure 4.

Figure 4 illustrates the calculated Seebeck coefficients for p-type SnSe in comparison with the measured data for doped SnSe [41–43,56,57]. The calculations were performed at the p-type doping levels of $4\times10^{19}$, $3\times10^{19}$, and $2\times10^{19}$ cm$^{-3}$ to compare with Hall data of $4\times10^{19}$ cm$^{-3}$. The selection of these theoretical doping levels was based on the observations that for the nominated carrier concentration of $4\times10^{19}$ cm$^{-3}$ by Zhao et al. [43], the measured carrier concentration has decreased from ~$4\times10^{19}$ cm$^{-3}$ to ~$2\times10^{19}$ cm$^{-3}$ at 800 K. Meanwhile, Zhao et al. [43] showed that for the measured Seebeck coefficients [41] with the nominated carrier concentration ~ $4\times10^{17}$ cm$^{-3}$, the real carrier concentration is ~$3\times10^{17}$ cm$^{-3}$ at 300 K. To compare with the measured data for the undoped SnSe [41] with the nominated carrier concentration of ~ $4\times10^{17}$ cm$^{-3}$, the calculations were preformed at the p-type doping levels of $3\times10^{17}$ and $4\times10^{17}$ cm$^{-3}$. The calculated mobile charge carrier concentrations for p-type SnSe are illustrated in Figure 5 and are in reasonable agreement with experimental data by Zhao el. [41].

## VI. CONCLUSION

In summary, we propose a first-principles thermodynamic procedure for calculating the Seebeck coefficient. It does not require the parabolic-band approximation or the value for the electron relaxation time. The formalism relies only on the e-DOS that is a routine output from most first-principles calculations in contrast to the commonly employed approach based on the Boltzmann transport theory that requires extremely dense reciprocal space mesh for the calculations of group velocity of electrons. Therefore, the present approach allows dramatically more efficient first-principles calculations of Seebeck coefficients than existing methods based on the Boltzmann transport theory, making it possible to search for superior thermoelectric materials based on routine first-principles calculations.

**Acknowledgements**

We would like to thank Prof. Jorge O. Sofo for providing the WIEN2k code, and Profs. Li-Dong Zhao, G. J. Snyder, Joseph Heremans, and David Singh for valuable discussions. This work was supported by the U.S. Department of Energy, Office of Basic Energy Sciences, Division of Materials Sciences and Engineering under Award DE-FG02-07ER46417 (Wang and Chen) and by National Science Foundation (NSF) through Grant Nos. DMR-1310289 and CHE-1230924 (Wang, Shang, Zhou, Hu, and Liu). First-principles calculations were carried out partially on the LION clusters at the Pennsylvania State University, partially on the resources of NERSC supported by the Office of Science of the U.S. Department of Energy under contract No. DE-AC02-05CH11231, and partially on the resources of XSEDE supported by NSF with Grant No. ACI-1053575.

[1] I. Vaskivskyi, J. Gospodaric, S. Brazovskii, D. Svetin, P. Sutar, E. Goreshnik, I.A. Mihailovic, T. Mertelj, and D. Mihailovic, Sci. Adv. **1**, e1500168 (2015).

[2] W.G. Zeier, J. Schmitt, G. Hautier, U. Aydemir, Z.M. Gibbs, C. Felser, and G.J. Snyder, Nat. Rev. Mater. **1**, 16032 (2016).

[3] C. Goupil, W. Seifert, K. Zabrocki, E. Müller, and G.J. Snyder, Entropy **13**, 1481 (2011).

[4] G.J. Snyder and E.S. Toberer, Nat. Mater. **7**, 105 (2008).

[5] K. Biswas, J. He, I.D. Blum, C.-I. Wu, T.P. Hogan, D.N. Seidman, V.P. Dravid, and M.G. Kanatzidis, Nature **489**, 414 (2012).

[6] Y. Pei, X. Shi, A. LaLonde, H. Wang, L. Chen, and G.J. Snyder, Nature **473**, 66 (2011).

[7] H.B. Callen, Phys. Rev. **73**, 1349 (1948).

[8] Y. Apertet, H. Ouerdane, C. Goupil, and P. Lecoeur, Eur. Phys. J. Plus **131**, 1 (2016).

[9] M. Cutler and N.F. Mott, Phys. Rev. **181**, 1336 (1969).

[10] T.J. Scheidemantel, C. Ambrosch-Draxl, T. Thonhauser, J. V Badding, and J.O. Sofo, Phys. Rev. B **68**, 125210 (2003).

[11] G.K.H. Madsen and D.J. Singh, Comput. Phys. Commun. **175**, 67 (2006).

[12] C. Wood, Reports Prog. Phys. **51**, 459 (1988).

[13] M.R. Peterson and B.S. Shastry, Phys. Rev. B **82**, 195105 (2010).

[14] J. Hejtmánek, Z. Jirák, and J. Šebek, Phys. Rev. B **92**, 125106 (2015).

[15] A.K. McMahan and M. Ross, Phys. Rev. B **15**, 718 (1977).

[16] N.D. Mermin, Phys. Rev. **137**, A1441 (1965).

[17] P. Giannozzi, S. Baroni, N. Bonini, M. Calandra, R. Car, C. Cavazzoni, D. Ceresoli, G.L. Chiarotti, M. Cococcioni, I. Dabo, A. Dal Corso, S. de Gironcoli, S. Fabris, G. Fratesi, R. Gebauer, U. Gerstmann, C. Gougoussis, A. Kokalj, M. Lazzeri, L. Martin-Samos, N. Marzari, F. Mauri, R. Mazzarello, S. Paolini, A. Pasquarello, L. Paulatto, C. Sbraccia, S. Scandolo, G. Sclauzero, A.P. Seitsonen, A. Smogunov, P. Umari, and R.M. Wentzcovitch, J. Phys. Condens. Matter **21**, 395502 (2009).

[18] X. Gonze, B. Amadon, P.-M. Anglade, J.-M. Beuken, F. Bottin, P. Boulanger, F. Bruneval, D. Caliste, R. Caracas, M. Côté, T. Deutsch, L. Genovese, P. Ghosez, M. Giantomassi, S. Goedecker, D.R. Hamann, P. Hermet, F. Jollet, G. Jomard, S. Leroux, M. Mancini, S. Mazevet, M.J.T. Oliveira, G. Onida, Y. Pouillon, T. Rangel, G.-M. Rignanese, D. Sangalli, R. Shaltaf, M. Torrent, M.J. Verstraete, G. Zerah, and J.W. Zwanziger, Comput. Phys. Commun. **180**, 2582 (2009).

[19] S.J. Clark, M.D. Segall, C.J. Pickard, P.J. Hasnip, M.I.J. Probert, K. Refson, and M.C. Payne, Zeitschrift Für Krist. - Cryst. Mater. **220**, 567 (2005).

[20] G. Kresse and J. Furthmüller, Comput. Mater. Sci. **6**, 15 (1996).

[21] G. Kresse and D. Joubert, Phys. Rev. B **59**, 1758 (1999).

[22] P. Blaha Karlheinz Schwarz Georg Madsen Dieter Kvasnicka Joachim Luitz, P. Blaha, K. Schwarz, G.K. H Madsen, D. Kvasnicka, J. Luitz, and U. Karlheinz Schwarz, (2001).

[23] J.M. Ziman, *Electrons and Phonons : The Theory of Transport Phenomena in Solids* (Clarendon Press, 2001).

[24] A. Sommerfeld, *Thermodynamics and Statistical Mechanics* (Academic Press, 1964).

[25] Y. Wang, J.J. Wang, H. Zhang, V.R. Manga, S.L. Shang, L.Q. Chen, and Z.K. Liu, J. Physics-Condensed Matter **22**, 225404 (2010).

[26] Y. Wang, D. Chen, and X. Zhang, Phys. Rev. Lett. **84**, 3220 (2000).

[27] P. Reddy, S.-Y. Jang, R.A. Segalman, and A. Majumdar, Science (80-. ). **315**, 1568 (2007).

[28] A. Jain, S.P. Ong, G. Hautier, W. Chen, W.D. Richards, S. Dacek, S. Cholia, D. Gunter, D. Skinner, G. Ceder, and K.A. Persson, APL Mater. **1**, 11002 (2013).

[29] S.P. Ong, W.D. Richards, A. Jain, G. Hautier, M. Kocher, S. Cholia, D. Gunter, V.L. Chevrier, K.A. Persson, and G. Ceder, Comput. Mater. Sci. **68**, 314 (2013).

[30] S. Bhattacharya and G.K.H. Madsen, Phys. Rev. B **92**, 85205 (2015).

[31] L.D. Hicks and M.S. Dresselhaus, Phys. Rev. B **47**, 12727 (1993).

[32] K. Behnia, *Fundamentals of Thermoelectricity* (Oxford University Press, 2015).

[33] D.K.C. MacDonald, *Thermoelectricity : An Introduction to the Principles* (John Wiley & Sons, Inc., New York, London, 1962).

[34] J.E. Parrott, IEEE Trans. Electron Devices **43**, 809 (1996).

[35] U. Sivan and Y. Imry, Phys. Rev. B **33**, 551 (1986).

[36] L. Gravier, S. Serrano-Guisan, F. Reuse, and J.-P. Ansermet, Phys. Rev. B **73**, 24419 (2006).

[37] M. Jonson and G.D. Mahan, Phys. Rev. B **21**, 4223 (1980).

[38] M. Jonson and G.D. Mahan, Phys. Rev. B **42**, 9350 (1990).

[39] J.P. Heremans, V. Jovovic, E.S. Toberer, A. Saramat, K. Kurosaki, A. Charoenphakdee, S. Yamanaka, and G.J. Snyder, Science (80-. ). **321**, 554 (2008).

[40] J.P. Heremans, C.M. Thrush, and D.T. Morelli, Phys. Rev. B **70**, 115334 (2004).

[41] L.-D. Zhao, S.-H. Lo, Y. Zhang, H. Sun, G. Tan, C. Uher, C. Wolverton, V.P. Dravid, and M.G. Kanatzidis, Nature **508**, 373 (2014).

[42] L.-D. Zhao, G. Tan, S. Hao, J. He, Y. Pei, H. Chi, H. Wang, S. Gong, H. Xu, V.P. Dravid, C. Uher, G.J. Snyder, C. Wolverton, and M.G. Kanatzidis, Science **351**, 141 (2016).

[43] L.-D. Zhao, C. Chang, G. Tan, M.G. Kanatzidis, X. Tan, Z. Liu, H. Qin, H. Shao, H. Jiang, B. Liang, J. Jiang, G.J. Snyder, C. Wolverton, M.G. Kanatzidis, and Z.F. Ren, Energy Environ. Sci. **9**, 3044 (2016).

[44] J.P. Perdew, K. Burke, and M. Ernzerhof, Phys. Rev. Lett. **77**, 3865 (1996).

[45] J.P. Perdew and A. Zunger, Phys. Rev. B **23**, 5048 (1981).

[46] D.J. Singh, Phys. Rev. B **81**, 195217 (2010).

[47] E. Engel and S.H. Vosko, Phys. Rev. B **47**, 13164 (1993).

[48] J.P. Perdew, A. Ruzsinszky, G.I. Csonka, O.A. Vydrov, G.E. Scuseria, L.A. Constantin, X. Zhou, and K. Burke, Phys. Rev. Lett. **100**, 136406 (2008).

[49] K. Adouby, C. Pérez-Vicente, J.C. Jumas, R. Fourcade, and A.A. Touré, Zeitschrift Für Krist. - Cryst. Mater. **213**, 343 (1998).

[50] Y. Wang, J.J. Wang, W.Y. Wang, Z.G. Mei, S.L. Shang, L.Q. Chen, and Z.K. Liu, J. Physics-Condensed Matter **22**, 202201 (2010).

[51] Z.K. Liu and Y. Wang, *Computational Thermodynamics of Materials* (Cambridge University Press, Cambridge, UK, UK, 2016).

[52] Y. Wang, S.L. Shang, Z.K. Liu, and L.Q. Chen, Phys. Rev. B **85**, 224303 (2012).

[53] Y. Wang, S.-L. Shang, H. Fang, Z.-K. Liu, and L.-Q. Chen, Npj Comput. Mater. **2**, 16006 (2016).

[54] M. Gajdoš, K. Hummer, G. Kresse, J. Furthmüller, and F. Bechstedt, Phys. Rev. B **73**, 45112 (2006).

[55] A.D. LaLonde, Y. Pei, G.J. Snyder, J. Zhang, B. Ren, Z. Wang, M.G. Kanatzidis, M.G. Kanatzidis, K.M. Paraskevopoulos, and M.G. Kanatzidis, Energy Environ. Sci. **4**, 2090 (2011).

[56] S. Sassi, C. Candolfi, J.-B. Vaney, V. Ohorodniichuk, P. Masschelein, A. Dauscher, and B. Lenoir, Appl. Phys. Lett. **104**, 212105 (2014).

[57] F. Serrano-Sánchez, M. Gharsallah, N.M. Nemes, F.J. Mompean, J.L. Martínez, and J.A. Alonso, Appl. Phys. Lett. **106**, 83902 (2015).

**Figure captions**

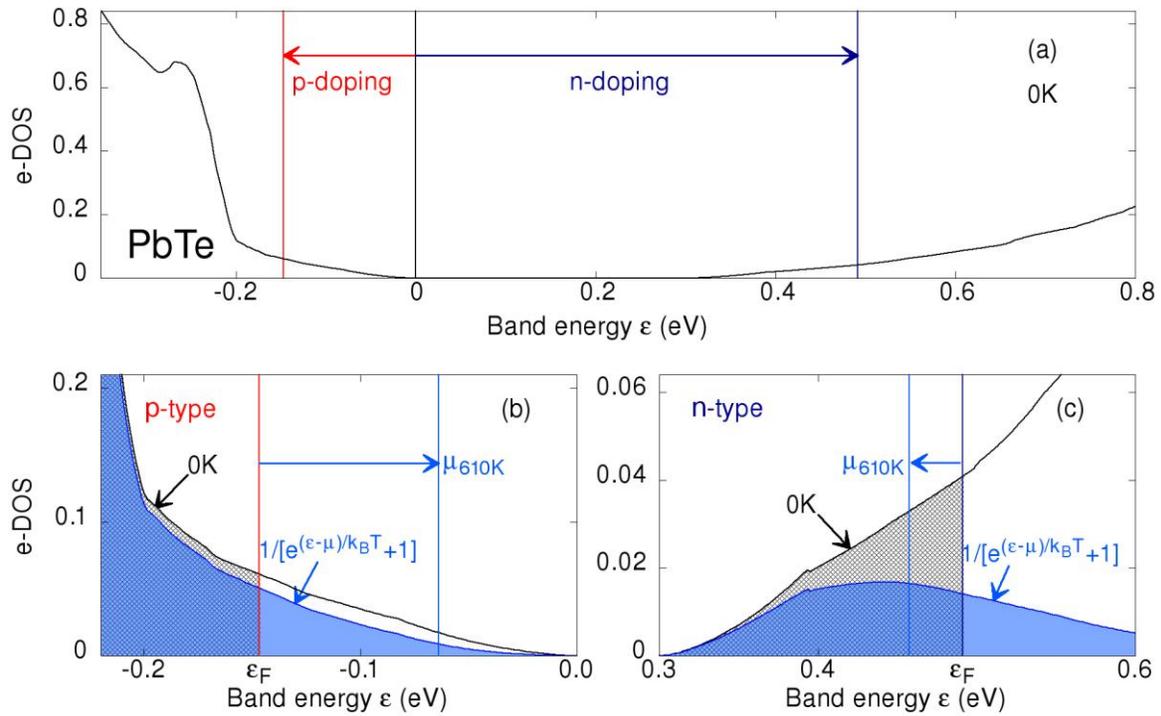

Figure 1. Illustration of the Seebeck effect: a) the calculated the electronic density of states (black curve); the p-doping shifts the Fermi energy towards lower energy as indicated by the red arrow pointed to the left; the n-doping increases the Fermi energy as indicated by the blue arrow pointed to the right; b) plot of electron density of states near the Fermi energy for p-doping; c) plot of electron density of near the Fermi energy for n-doping. For b) and c), the areas shaded by gray (partially overlapped by the blue semitransparent shaded areas) represent the electron occupation at 0 K and the blue semitransparent shaded areas indicate the electron occupation at a finite temperature 610 K described by the Fermi distribution.

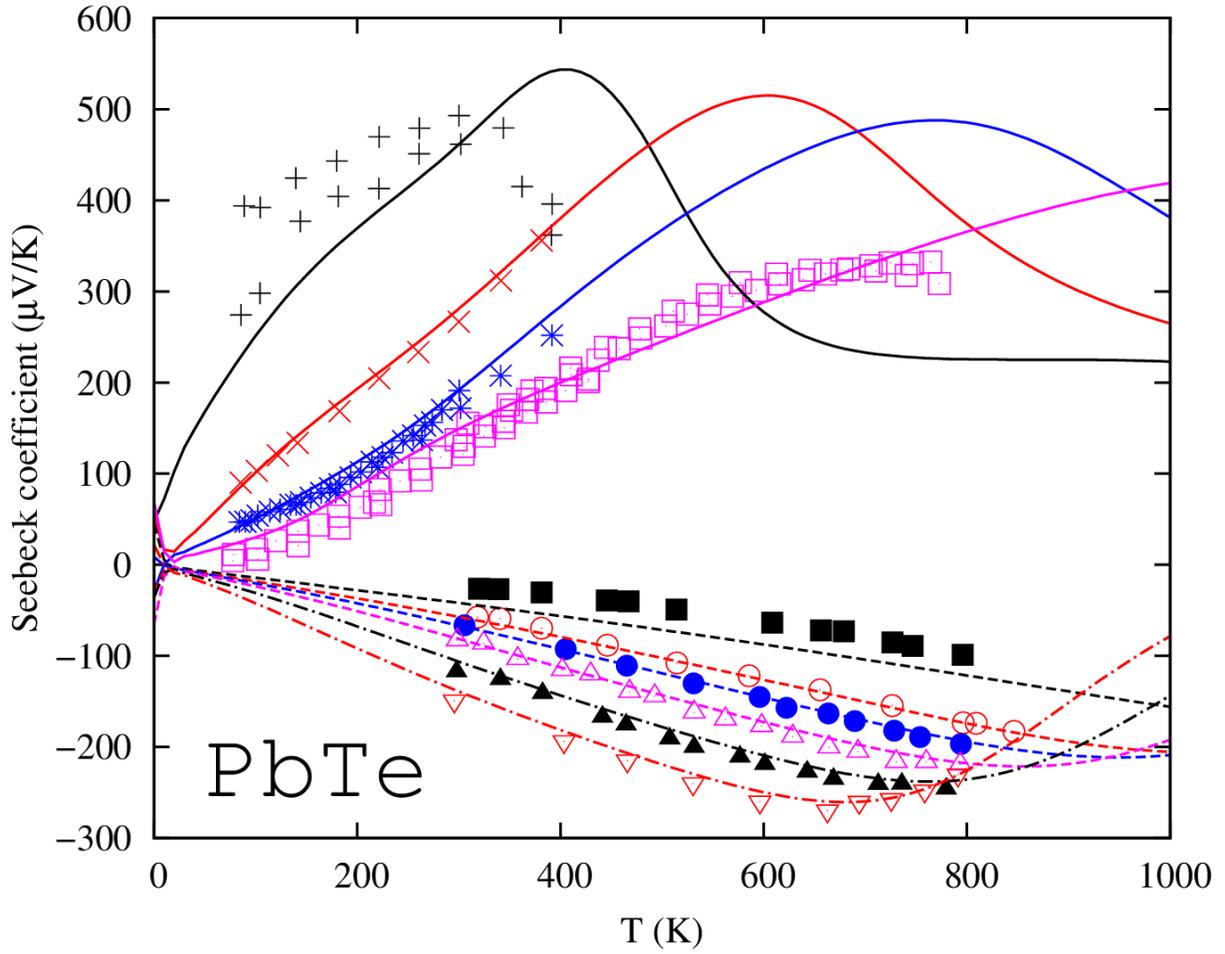

| Heremans,p=2.0e17 | + | This work,p=2.0e17 | ——— |
| Heremans,p=2.0e18 | × | This work,p=2.0e18 | ——— |
| Heremans,p=8.0e18 | ✳ | This work,p=8.0e18 | ——— |
| Heremans,p=5.3e19 | □ | This work,p=5.3e19 | ——— |
| LaLonde,n=1.4e20 | ■ | This work,n=1.4e20 | - - - |
| LaLonde,n=4.0e19 | ○ | This work,n=4.0e19 | - - - |
| LaLonde,n=2.8e19 | ● | This work,n=2.8e19 | - - - |
| LaLonde,n=1.7e19 | △ | This work,n=1.7e19 | - - - |
| LaLonde,n=1.0e19 | ▲ | This work,n=1.0e19 | -·-·- |
| LaLonde,n=5.8e18 | ▽ | This work,n=5.8e18 | -·-·- |

Figure 2. The calculated Seebeck coefficients for PbTe for p-type doping levels of $2.0\times10^{17}$, $2.0\times10^{18}$, $8.0\times10^{18}$, and $5.3\times10^{19}$ cm$^{-3}$ and n-type doping levels of $1.4\times10^{20}$, $4.0\times10^{19}$, $2.8\times10^{19}$, $1.7\times10^{19}$, $1.0\times10^{19}$, and $5.8\times10^{18}$ cm$^{-3}$. Corresponding to these number sequence, the lines

represent the present calculations while the symbols (with same colors and sequences as the lines) represent the measured data for p-type PbTe by Heremans and coauthors [39,40] and n-type PbTe by LaLonde et al. [55]

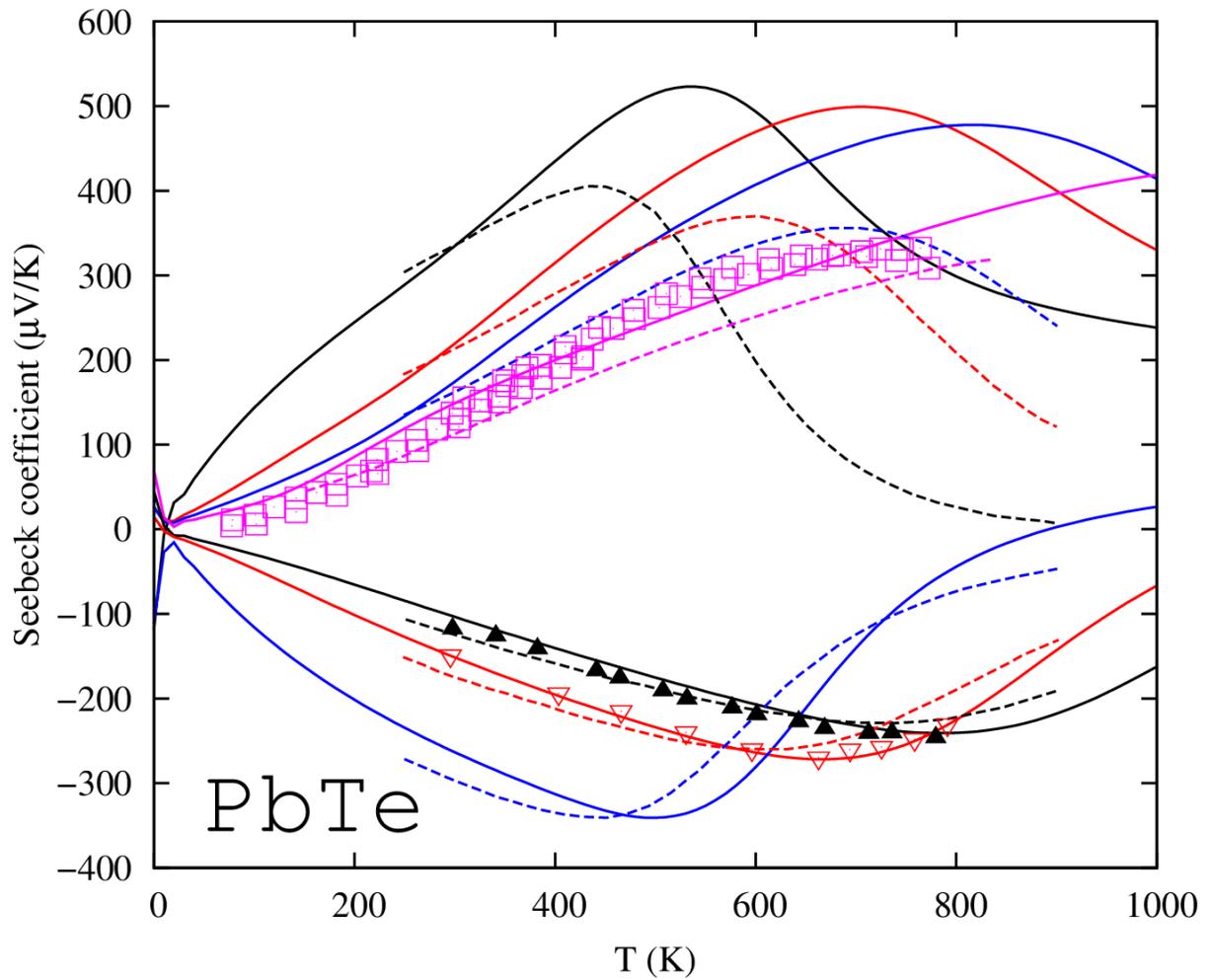
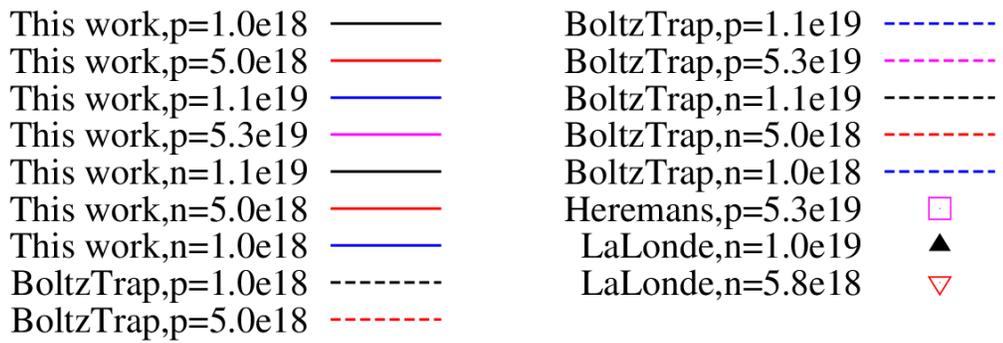

Figure 3. Comparison between the present approach and Boltzmann transport theory. The dots represent the measured data at p-type doping levels of $5.3\times10^{19}$ cm$^{-3}$ by Heremans and coauthors [21,22] and n-type doping levels of $1.0\times10^{19}$ and $5.8\times10^{18}$ cm$^{-3}$ by LaLonde et al. [55].

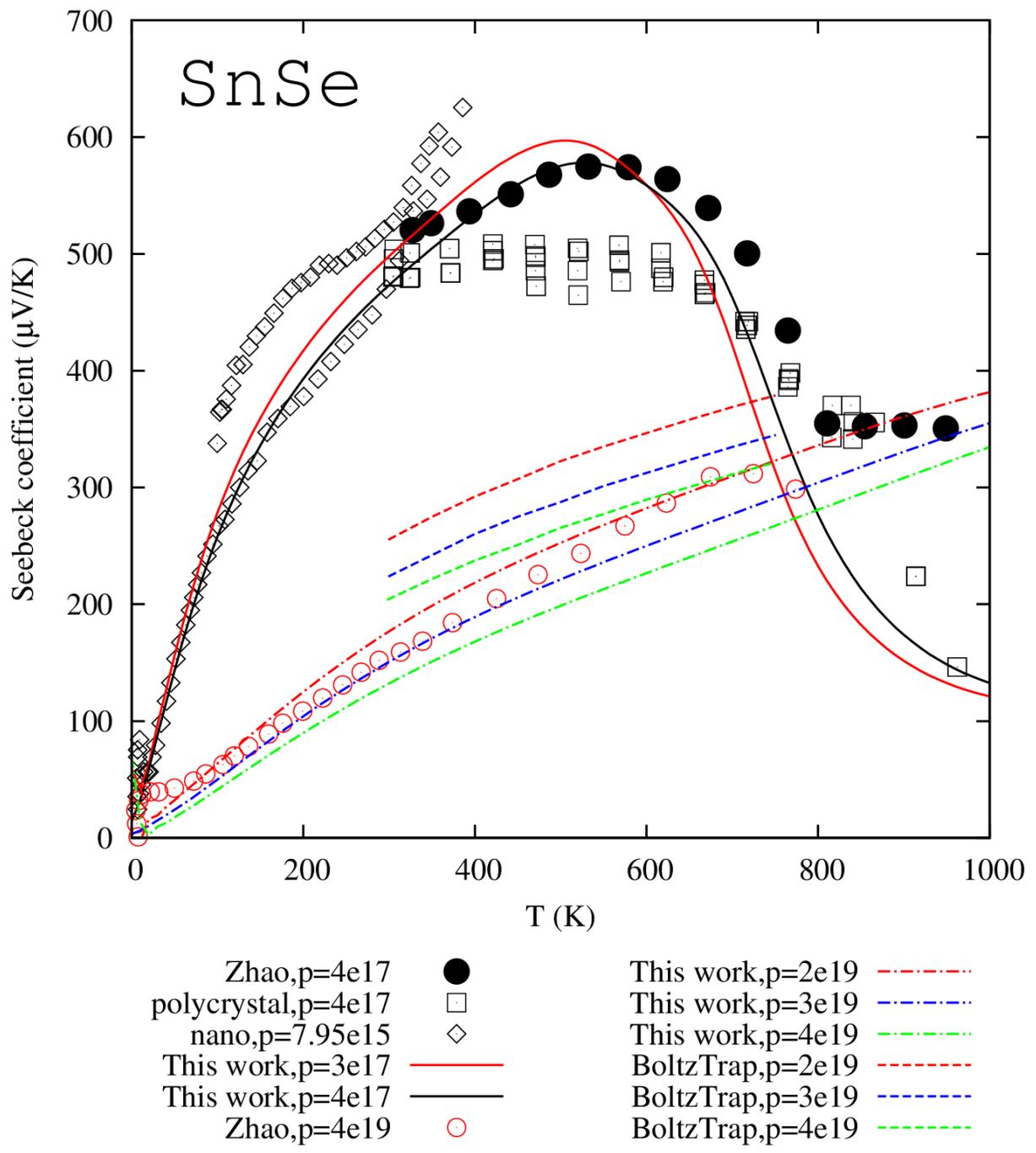

Figure 4. The calculated Seebeck coefficients for p-type SnSe compared with experiments [41,42,56,57].

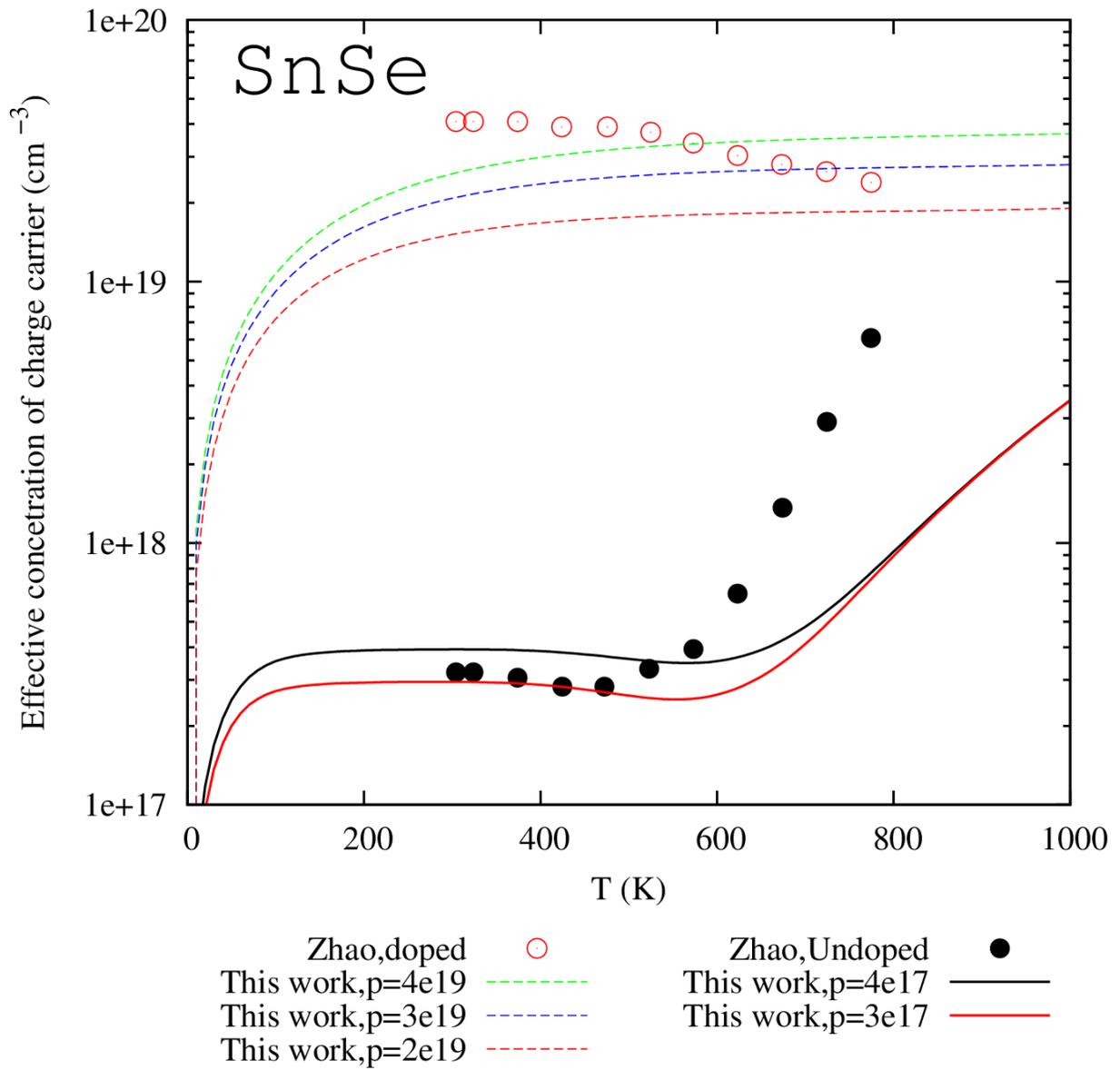

Figure 5. The mobile charge carrier concentrations for p-type SnSe. The open and closed circles represent experimental data for p-type doping with nominated carrier concentrations, respectively, of ~4×10$^{19}$ cm$^{-3}$ and ~4×10$^{17}$ cm$^{-3}$, from Zhao et al. [41]